\PassOptionsToPackage{table}{xcolor}

\documentclass[manuscript,sigconf]{acmart}
\AtBeginDocument{%
  \providecommand\BibTeX{{%
    \normalfont B\kern-0.5em{\scshape i\kern-0.25em b}\kern-0.8em\TeX}}}

\setcopyright{acmcopyright}
\copyrightyear{2023}
\acmYear{2023}
\setcopyright{rightsretained}
\acmConference[MobileHCI '23 Companion]{25th International Conference on Mobile Human-Computer Interaction}{September 26--29, 2023}{Athens, Greece}
\acmBooktitle{25th International Conference on Mobile Human-Computer Interaction (MobileHCI '23 Companion), September 26--29, 2023, Athens, Greece}\acmDOI{10.1145/3565066.3608685}
\acmISBN{978-1-4503-9924-1/23/09}

\usepackage{subcaption}
\usepackage{color,soul,bm}

\newcommand{\rev}[1]{\textcolor{black}{#1}}

\usepackage{tikz}
\usetikzlibrary{backgrounds}

\begin{document}

\title[The State of Algorithmic Fairness in Mobile Human-Computer Interaction]{The State of Algorithmic Fairness in Mobile Human-Computer Interaction}

\author{Sofia Yfantidou}
\email{syfantid@csd.auth.gr}
\orcid{0000-0002-5629-3493}
\affiliation{%
  \institution{Aristotle University of Thessaloniki}
  \city{Thessaloniki}
  \country{Greece}
}

\author{Marios Constantinides}
\authornotemark[1]
\email{marios.constantinides}
\email{@nokia-bell-labs.com}
\orcid{0000-0003-1454-0641}
\affiliation{%
  \institution{Nokia Bell Labs}
  \city{Cambridge}
  \country{United Kingdom}
}

\author{Dimitris Spathis}
\authornote{Also affiliated with the University of Cambridge, UK.}
\email{dimitrios.spathis}
\orcid{0000-0001-9761-951X}
\email{@nokia-bell-labs.com}
\affiliation{%
  \institution{Nokia Bell Labs}
  \city{Cambridge}
  \country{United Kingdom}
}

\author{Athena Vakali}
\email{avakali@csd.auth.gr}
\orcid{0000-0002-0666-6984}
\affiliation{%
  \institution{Aristotle University of Thessaloniki}
  \city{Thessaloniki}
  \country{Greece}
}

\author{Daniele Quercia}
\email{daniele.quercia@nokia-bell-labs.com}
\orcid{0000-0001-9461-5804}
\affiliation{%
  \institution{Nokia Bell Labs}
  \city{Cambridge}
  \country{United Kingdom}
}

\author{Fahim Kawsar}
\email{fahim.kawsar@nokia-bell-labs.com}
\orcid{0000-0001-5057-9557}
\affiliation{%
  \institution{Nokia Bell Labs}
  \city{Cambridge}
  \country{United Kingdom}
}

\renewcommand{\shortauthors}{Yfantidou, et al.}

\begin{abstract}

This paper explores the intersection of \rev{Artificial Intelligence and Machine Learning (AI/ML)} fairness and mobile human-computer interaction (MobileHCI). Through a comprehensive analysis of MobileHCI proceedings published between 2017 and 2022, we first aim to understand the current state of algorithmic fairness in the community. By manually analyzing 90 papers, we found that only a small portion (5\%) thereof adheres to modern fairness reporting, such as analyses conditioned on demographic breakdowns. At the same time, the overwhelming majority draws its findings from highly-educated, employed, and Western populations. We situate these findings within recent efforts to capture the current state of algorithmic fairness in mobile and wearable computing, and envision that our results will serve as an open invitation to the design and development of fairer ubiquitous technologies.

\end{abstract}



\maketitle

\section{Introduction}
\label{sec:introduction}

The rapid growth of mobile and wearable technology, along with its widespread integration into our daily lives, has raised concerns about the ethical implications of algorithmic decision-making~\cite{shneiderman2021responsible, kissinger2021age, tahaei2023human}. Algorithms deployed on mobile and wearable devices are currently used to recognize human activities~\cite{gu2021survey}, detect breathing~\cite{10.1145/3369835}, and infer sleep quality~\cite{koskimaki2018we}. However, numerous reports have highlighted instances where \rev{Artificial Intelligence (AI) algorithms} exhibit bias. For example, image classification systems consistently misclassify individuals of color~\cite{buolamwini2018gender}, and health sensors like oximeters have shown bias due to primarily testing on white participants~\cite{sjoding2020racial}. These examples necessitate a call to action, emphasizing the need to prioritize the design and development of fair mobile and wearable systems \cite{yfantidouworkshop}. In practical terms, this means shifting focus from \emph{algorithmic performance} to \emph{algorithmic fairness}.

To achieve that, we first need to map out the space of algorithmic fairness in mobile and wearable computing. By utilizing a methodology in line with ours, Yfantidou et al.~\cite{yfantidou2023beyond} captured the current state of algorithmic fairness in adjacent venues to MobileHCI such as the ACM IMWUT, MobiCom, MobiSys, and SenSys, IEEE Pervasive, and Transactions on Mobile Computing Computing. Along these lines, \citet{linxen2021weird} conducted a meta-study on CHI proceedings from 2016 to 2020, reporting that 73\% of those papers are based on Western populations. A similar picture was also found at the ACM FAccT proceedings~\cite{ali_weird_2023}; a scientific community that is highly concerned with fairness, accountability, and transparency in AI systems. More broadly, addressing algorithmic bias in \rev{Machine Learning (ML)} has been a longstanding issue, despite its recent surge. For example, \citet{caton2020fairness} and \citet{pessach2022review} reviewed fairness metrics and bias mitigation approaches, while \citet{le2022survey} surveyed available datasets for fairness research, including financial, healthcare, social, and educational datasets some of which might well be applicable to MobileHCI research. \rev{More targeted, \citet{olteanu2019social} reviewed the literature surrounding social data biases, such as biases in user-generated content, expressed or implicit relations between people, and behavioral traces. On a different note, \citet{10.1145/3173574.3174156} featured emerging trends for explainable, accountable, and intelligible systems within the CHI community, also discussing notions of fairness.} While these surveys and reports provide insight into the state of algorithmic fairness in areas related to MobileHCI, we still lack a comprehensive understanding of the current situation within this specific community. Therefore, we seek to find answers to how fair is MobileHCI, and, in doing so, answer our main research questions: \emph{What is the current state of algorithmic fairness in MobileHCI?}

\begin{figure}[htb!]
  \centering
  \includegraphics[width=.7\linewidth]{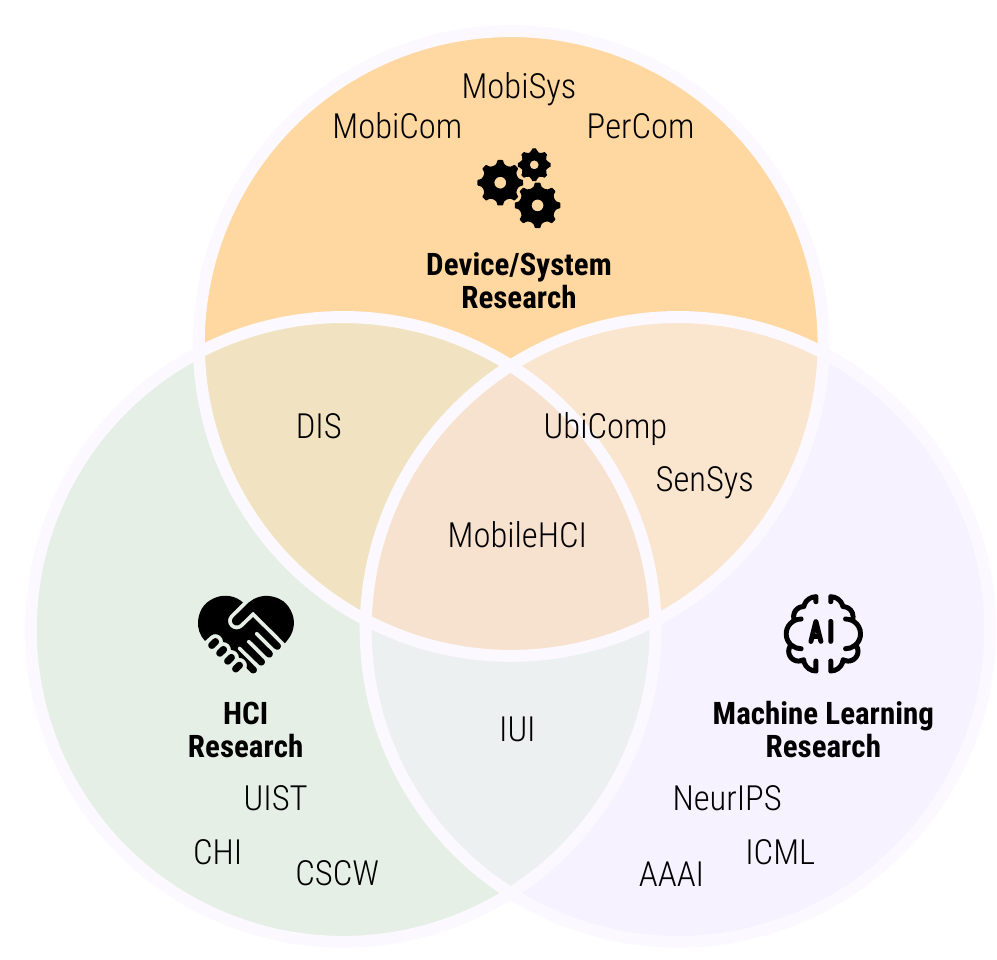}
  \caption{\textbf{MobileHCI sits on the intersection of ML, HCI, and system and device research.} While fairness has been discussed independently in each domain, the synergy among them, as captured by MobileHCI remains unexplored. Note that the conference categorization is subjective and based on the authors' experiences and interpretations.\label{fig:venn_diagram}}
  \Description{A Venn diagram with three partially overlapping circles representing Device/System Research, HCI Research and ML Research. Publications venues are placed in the circles depending on their relevance with the aforementioned research scopes. MobiSys, MobiCom, PerCom fall under Device/System Research, UIST, CHI, and CSCW fall under HCI Research, DIS in between the two, NeurIPS, AAAI, and ICML fall under ML Research, and IUI between the latter two. Finally UbiComp and SenSys fall under ML and Device/System Research, whereas MobileHCI lies at the heart of all three.}
\end{figure}
By answering this question, we foresee continuing the discussion about algorithmic fairness in mobile and wearable computing by bringing the human-centred aspect of the MobileHCI community. Establishing itself at the crossovers between device/system research, ML research, and human-computer interaction (HCI) (Figure~\ref{fig:venn_diagram}), we believe MobileHCI is well-positioned to be a champion for fair interaction(s) with and through mobile devices, applications, and services.  We made three sets of contributions: 

\begin{itemize}
    \item We conducted a literature review of algorithmic fairness in MobileHCI proceedings between 2017-2022, where we screened 90 papers and critically reviewed \textbf{14} of them (\S\ref{sec:methodology}). 
    \item We found that only \textbf{5\%} of all MobileHCI papers published between 2017-2022 adhere to modern fairness reporting, such as analyses conditioned on demographic breakdowns, while the overwhelming majority draws its findings from highly-educated, employed, and Western populations (\S\ref{sec:results}).
    \item By situating MobileHCI in the broader mobile and wearable computing research, we discuss how the community needs to move forward and learn from adjacent communities (\S\ref{sec:discussion}).
\end{itemize}

\section{Methodology\label{sec:methodology}}

\begin{figure}[]
          \centering
          \includegraphics[width=\linewidth]{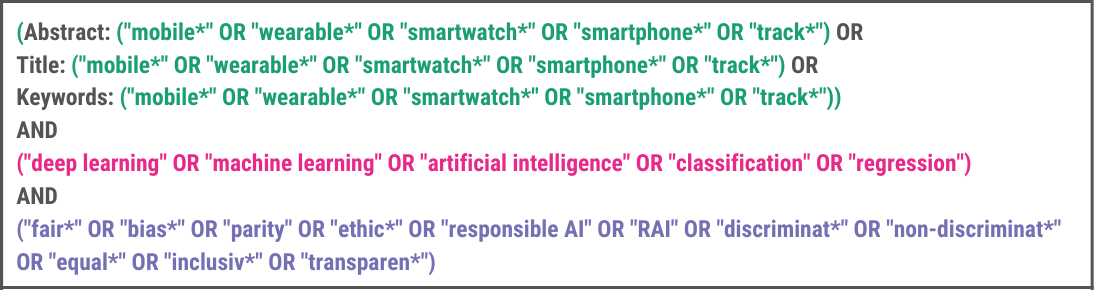}
          \caption{\textbf{The query utilized for recovering relevant papers from the ACM Digital Library}. Terms related to UbiComp are highlighted in green, ML in pink, and fairness in purple.\label{fig:query}}
          \Description{The query utilized for the paper retrieval that consists of three parts: first, we looks for words, such as mobile, wearable, or smartwatch and tracker in the meta-data of the paper; second, we look for words such as deep learning, machine learning and artificial intelligence in the full-text and meta-data; and third, same for the words related to bias, fairness, ethics, and responsible AI.}
\end{figure}

In this work, we replicate the methodology employed by \citet{yfantidou2023beyond} to maintain consistency in the approach and enhance the comparability of results. \rev{Specifically,} we started by conducting a comprehensive screening of relevant papers published in MobileHCI between 2017 and 2022 \rev{to capture emerging trends in fairness and MobileHCI research}. \rev{The search query was formulated using the original keywords (Figure~\ref{fig:query}) to ensure a comprehensive search and facilitate cross-community comparisons. Prior work \cite{yfantidou2023beyond} has attempted to extend the query keywords without retrieving additional publications. Nevertheless, we recognize certain methodological limitations related to query definition. For instance, a paper that does not explicitly utilize the terms AI or ML could be unfairly categorized as not addressing fairness. However, the query definition for the study incorporated terminology used in relevant review papers on fairness \cite{le2022survey,caton2020fairness}. Additionally, according to Fjeld et al.’s meta-analysis of prominent AI principles documents \cite{fjeld2020principled}, every document was referencing at least one of the following principles: ``non-discrimination and the prevention of
bias'', ``representative and high-quality data'', ``fairness'', ``equality'', ``inclusiveness in impact'', and ``inclusiveness in design'', mostly included in our query’s coverage. The query was further refined by consulting Responsible Artificial Intelligence (RAI) white papers from major tech companies. Finally,} the eligibility assessment was performed by three independent reviewers, while Kitchenaham and Charters' protocol \cite{kitchenham2007guidelines} was followed to ensure a systematic and transparent approach throughout the review process. \rev{To minimize the risk of bias in the included studies, the three reviewers conducted their assessments independently, and any differences or conflicts were resolved through discussion and consensus. No automation tools were used in the process.} By replicating this methodology, we aimed to provide a reliable and rigorous analysis of the literature in our field of study. 
To ensure the high quality and relevance of the included papers, we defined appropriate exclusion criteria that helped us determine the included papers:
\begin{enumerate}
    \item Papers lacking a quantitative evaluation of empirical or artefact contributions in mobile and wearable computing;
    \item Papers failing to include a quantitative assessment of bias or performance discrepancies in their evaluation, specifically concerning sensitive attributes like age, gender, race, disability, religion, sexual orientation, and more;
    \item Papers discussing different domains, such as computer vision, without integrating a ubiquitous component;
    \item Papers referencing bias in different contexts, such as the bias-variance trade-off or the bias parameter in neural networks, rather than addressing bias within the intended scope. 
\end{enumerate}

\begin{figure}[]
          \centering
          \includegraphics[width=\linewidth]{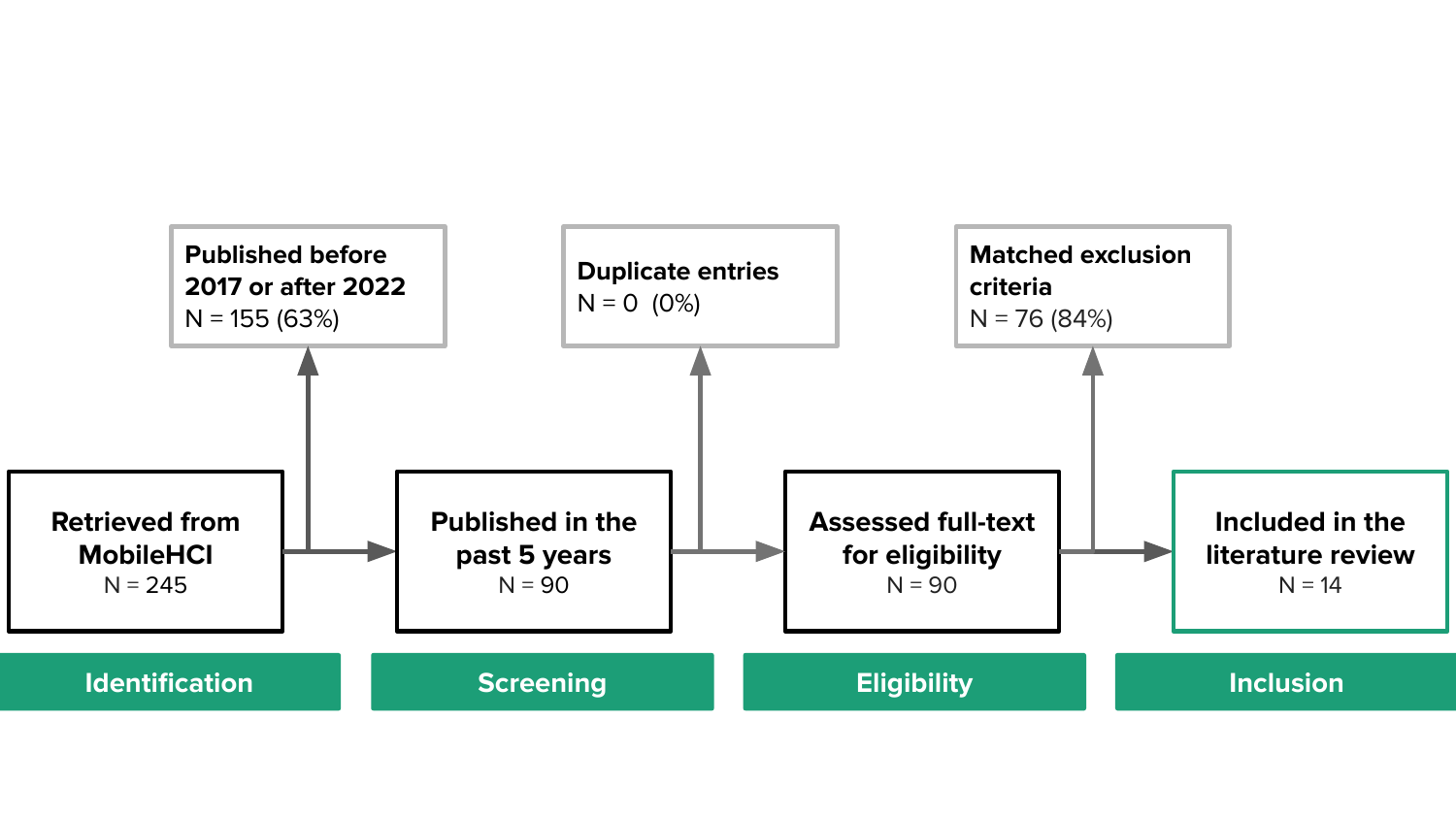}
          \caption{\textbf{PRISMA diagram}. Out of the 90 retrieved papers, only $\sim$16\% did not check any exclusion criterion.  \label{fig:prisma}}
          \Description{The PRISMA diagram of the literature review process. Out of the 245 papers retrieved by the search query, 90 were published within the past 5 years and were assessed for full-text eligibility, and 14 were included in the literature review, based on the exclusion criteria.}
\end{figure}

As seen in the Preferred Reporting Items for Systematic Reviews and Meta-Analyses (PRISMA) \cite{moher2009preferred} flow diagram of Figure~\ref{fig:prisma}, we screened 90 papers after date filtering and duplicate elimination. We then excluded 76 based on the pre-defined exclusion criteria. Hence, we included 14 papers in this review. \rev{Regarding paper synthesis, the coded variables and the full codebook can be found on GitHub for transparency \cite{sofia_yfantidou_2023_8143676}}.

\section{Results}
\label{sec:results}
We first discuss the state of fairness in MobileHCI (\S\ref{stateFairness}), followed by alternative notions of fairness that this specific community has indirectly adopted (\S\ref{AlternativeNotions}) and the diversity in study design (\S\ref{WEIRDSection}).

\subsection{What is the State of Fairness in MobileHCI?\label{stateFairness}}

\renewcommand{\arraystretch}{1.1}
\begin{table*}[]
\caption{\textbf{Included papers categorized by application domain, deployment setting, and protected attribute.} Qualitative and quantitative studies on understanding people are the most active in the MobilHCI community regarding fairness considerations. }
\label{tab:summary}
\begin{tabular}{llll}
\rowcolor[HTML]{1C9E77} 
{\color[HTML]{FFFFFF} \textbf{APPLICATION DOMAIN}} & {\color[HTML]{FFFFFF} \textbf{DEPLOYMENT SETTING}} & {\color[HTML]{FFFFFF} \textbf{PROTECTED ATTRIBUTE}} & {\color[HTML]{FFFFFF} \textbf{PAPERS}}                 \\
\rowcolor[HTML]{F3F3F3} 
Understanding People (23.1\%)                      & In-the-wild                                        & Age, Gender                                         & \cite{10.1145/3338286.3340120,10.1145/3229434.3229441} \\
\rowcolor[HTML]{F3F3F3} 
                                                   &                                                    & Nationality                                         & \cite{10.1145/3338286.3340120,10.1145/3229434.3229474} \\
\rowcolor[HTML]{F3F3F3} 
                                                   &                                                    & Socioeconomic Status                                & \cite{10.1145/3229434.3229474}                         \\
User Experience and Usability (23.1\%)             & In-the-wild                                        & Age                                                 & \cite{10.1145/3098279.3098532}                         \\
                                                   &                                                    & Gender                                              & \cite{10.1145/3098279.3098532,10.1145/3098279.3098549} \\ 
                                                   & In-the-lab                                         & Nationality                                         & \cite{10.1145/3098279.3098555}                          \\
\rowcolor[HTML]{F3F3F3} 
Interaction beyond the Individual (15.4\%)         & In-the-wild                                        & Gender                                              & \cite{10.1145/3098279.3098559,10.1145/3098279.3098551} \\
\rowcolor[HTML]{F3F3F3} 
                                                   &                                                    & Miscellaneous                                       & \cite{10.1145/3098279.3098559}                         \\
Interacting with Devices (15.4\%)                  & In-the-wild                                        & Age, Gender, Nationality                & \cite{10.1145/3447526.3472059}                         \\ 
                                                   & In-the-lab                                         & Physiology                                          & \cite{10.1145/3338286.3340115,10.1145/3447526.3472059}                         \\
\rowcolor[HTML]{F3F3F3} 
Health (7.7\%)                                     & Both                                               & Health Condition                                    & \cite{10.1145/3379503.3403543}                         \\
Accessibility \& Ageing (7.7\%)                     & In-the-lab                                         & Age                                                 & \cite{10.1145/3338286.3340121}                         \\
\rowcolor[HTML]{F3F3F3} 
Specific Application Areas (7.7\%)                 & In-the-wild                                        & Age, Gender                                         & \cite{10.1145/3098279.3098552}                        
\end{tabular}
\end{table*}

Table~\ref{tab:summary} presents the included papers grouped by application domain, deployment setting, and protected attributes. In summary, out of 261 papers published in MobileHCI between 2017 and 2022 ($N_{all}=261$), only a small fraction of 5\% matched the screening criteria and were included for review ($N_{included}=14$), highlighting the timeliness of this work. 
\smallskip 

\noindent\textbf{Application Domain.} To identify appropriate application domains, we adopted CHI's subcommittees, identifying seven themes: Understanding People, User Experience and Usability, Interaction beyond the Individual, Interacting with Devices, Health, Accessibility \& Ageing, and Specific Application Areas, presented in descending order of frequency.
\smallskip

\noindent\textbf{Deployment Setting.} While laboratory settings offer controlled environments that facilitate experimentation (i.e., in-the-lab studies), they may not fully capture the intricacies of the real world where the applications are deployed. Consequently, there is a growing trend toward conducting in-the-wild studies, focusing on evaluating the situated design experience of MobileHCI. 
Contrary to publication venues soliciting primarily artefact or early-stage work papers \cite{yfantidou2023beyond}, in-the-wild studies prevail in MobileHCI ($\sim70\%$ of included papers), followed by in-the-lab studies (23\%) while a small fraction of papers ($\sim$8\%) report results for both deployments. Interestingly, biases were encountered across all deployment settings. \rev{Specifically, 22\% of in-the-wild and 33\% of in-the-lab studies reported bias against certain protected attributes, whereas all studies considering both deployments also encountered some form of bias.}
\smallskip

\noindent\textbf{Protected Attributes and Input Modalities.} In accordance with the Charter of Fundamental Rights \cite{eu2012charter} and previous research on fairness \cite{pessach2022review}, MobileHCI studies explored a range of protected attributes, as depicted in Table~\ref{tab:summary}. The representation of these attributes varied among the included papers, with gender and age being the most frequently mentioned (appearing in 9 and 6 out of 14 papers, respectively), followed by nationality (4 papers), physiology (2 papers), health conditions and socioeconomic status (1 paper). In line with prior work on UbiComp fairness \cite{yfantidou2023beyond}, attributes with a history of discrimination in machine learning, such as race \cite{compas,manzini2019black}, were not addressed in the studied literature. However, neither UbiComp nor HCI research are exempt from such disparities. Recent studies by \citet{sjoding2020racial} uncovered racial and ethnic biases in pulse oximetry, while \citet{10.1145/3411764.3445488} showed that 73\% of CHI study findings are based on Western participant samples. Additionally, MobileHCI (as captured by the included papers) uses distinct modalities compared to other UbiComp communities. For instance, the primary modalities evaluated for bias in the reviewed papers were usage/interaction logs (57\% of included papers) and self-reports (50\%). In contrast, sensor data, image, video, and audio data are more prevalent in ubiquitous machine learning and system and device research \cite{yfantidou2023beyond}. These findings underscore the importance of evaluating our hypotheses and models on diverse populations across multiple relevant attributes and input modalities. 
\smallskip

\noindent\textbf{Biases.} In the included papers, we encountered biases towards females and users with glasses in crowdsourcing for mobile eye tracking, as \textit{``workers  responses  were less accurate  with data  of  participants  wearing  spectacles and eye make-up''} \cite{10.1145/3098279.3098559}. This was mainly due to the partial occlusion of the pupil from  the  eye  camera  field  of  view caused by the eyewear's frame (such as spectacles or contact lenses) or the presence of eye makeup, introducing noise that hindered [the model's] utility. Similarly, we found biases toward users with respiratory conditions on models estimating lung function, where \citet{10.1145/3379503.3403543} ``observed that the mean absolute error for healthy participants is lower than other [Asthma, and COPD] patients''. Finally, interaction studies were not immune to biases, as physiological features, such as hand and thumb size, affected touch accuracy, where \textit{``[...] the length of the thumb is the most influential feature from [the] set. Intuitively, this makes sense as thumb length has a direct impact on what parts of the screen are reachable''} \cite{10.1145/3338286.3340115}. \rev{Note that biases are defined as systematic performance discrepancies across groups, as reported by the authors.} Yet despite the encountered biases, we found no usages of bias mitigation approaches or fairness metrics in the included papers.

\begin{figure}[]
    \centering
    \begin{subfigure}[t]{0.45\columnwidth} %
        \centering
        \includegraphics[width=\linewidth]{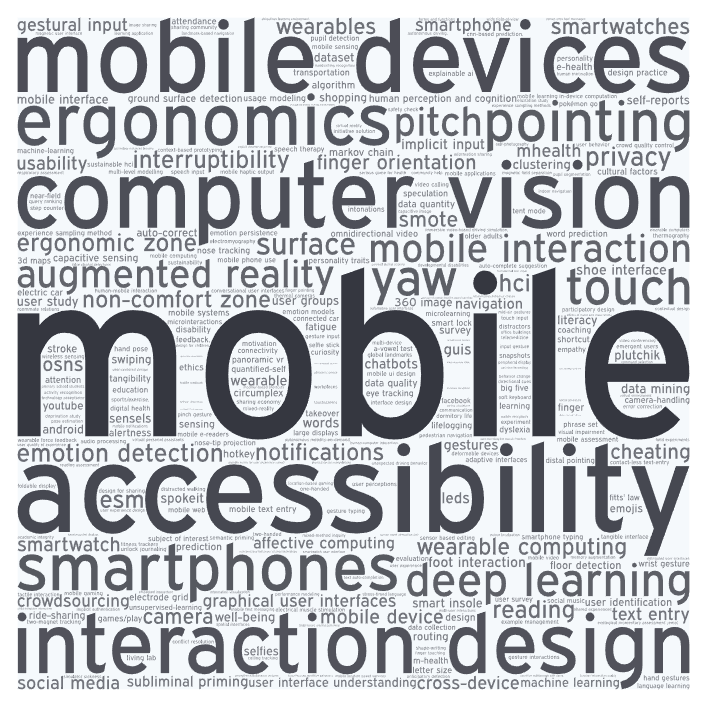}
    \end{subfigure}%
    \hspace{1em}%
    \begin{subfigure}[t]{0.45\columnwidth} %
        \centering
        \includegraphics[width=\linewidth]{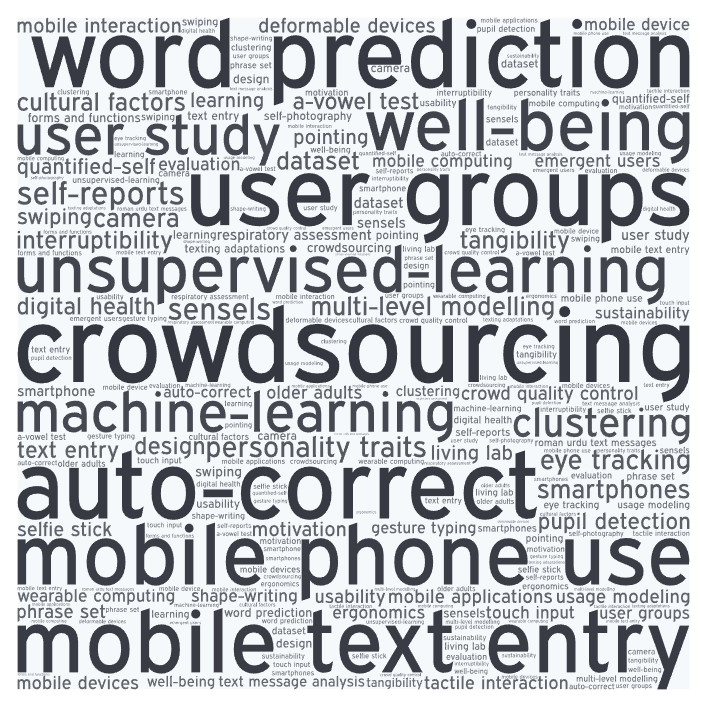}
    \end{subfigure}
    \caption{\textbf{Keyword differences of retrieved (left) and included (right) papers}. Interaction and design factors dominate the keywords in the included papers. \label{fig:wordclouds}}
    \Description{Word clouds representing the keyword distribution in the retrieved and included papers. In the retrieved papers, keywords such as mobile, mobile devices, ergonomics, computer vision, pointing, accessibility, smartphones, interaction design, and deep learning prevail. In the included papers, we see different prevalent keywords, such as crowdsourcing, user groups, word prediction, well-being, user study, auto-correct, mobile phone use, mobile text entry, unsupervised learning and machine learning.}
\end{figure}

\subsection{Alternative notions of fairness in MobileHCI\label{AlternativeNotions}}
\noindent\textbf{Diverse Samples \& Data Exploration.} 
Due to the nature of the community's work, we did not exclude papers conducting exploratory analysis conditioned on protected attributes, regardless of whether they incorporated ML techniques (50\% of the papers examined). Among these papers, we discovered a higher diversity of samples, with 36\% of them featuring US-based samples. This percentage is significantly lower than the 86\% that applies to other UbiComp subcommunities \cite{yfantidou2023beyond}. Notably, none of the included papers shared the exact same recruitment countries, indicating a global perspective in research efforts. Despite the initial prevalence of WEIRD samples \cite{linxen2021weird}, our findings suggest a promising trend toward democratizing HCI within the mobile computing community.\\

\noindent\textbf{Application Domains \& Ablation Studies.} 
The most prominent domains \rev{assigned by the authors of this work and observed} within the papers included in our study were ``Understanding People'' and ``User Experience and Usability'', followed by ``Interaction Beyond the Individual'' and ``Interacting with Devices''. This trend is further supported by Figure~\ref{fig:wordclouds}, where keywords such as ``crowdsourcing'', ``mobile phone use'', and ``mobile text entry'' emerge as the most frequent terms in the included papers (right), in contrast to the retrieved papers (left). Notably, these domains differ from prevalent research areas in other UbiComp subcommunities, where ``Health'', ``Privacy and Security'', and ``Human-Activity Recognition'' dominate \cite{yfantidou2023beyond}. This disparity in research focus also translates into the types of ablation studies within the MobileHCI community. Rather than emphasizing environmental conditions and other external factors, these studies primarily explore interaction-related factors, such as swipe finger and screen size. Nevertheless, the community strives for generalizability through rigorous ablation studies, ensuring the robustness and applicability of their models and findings.
\begin{figure*}[]
  \centering
  \includegraphics[width=.7\linewidth]{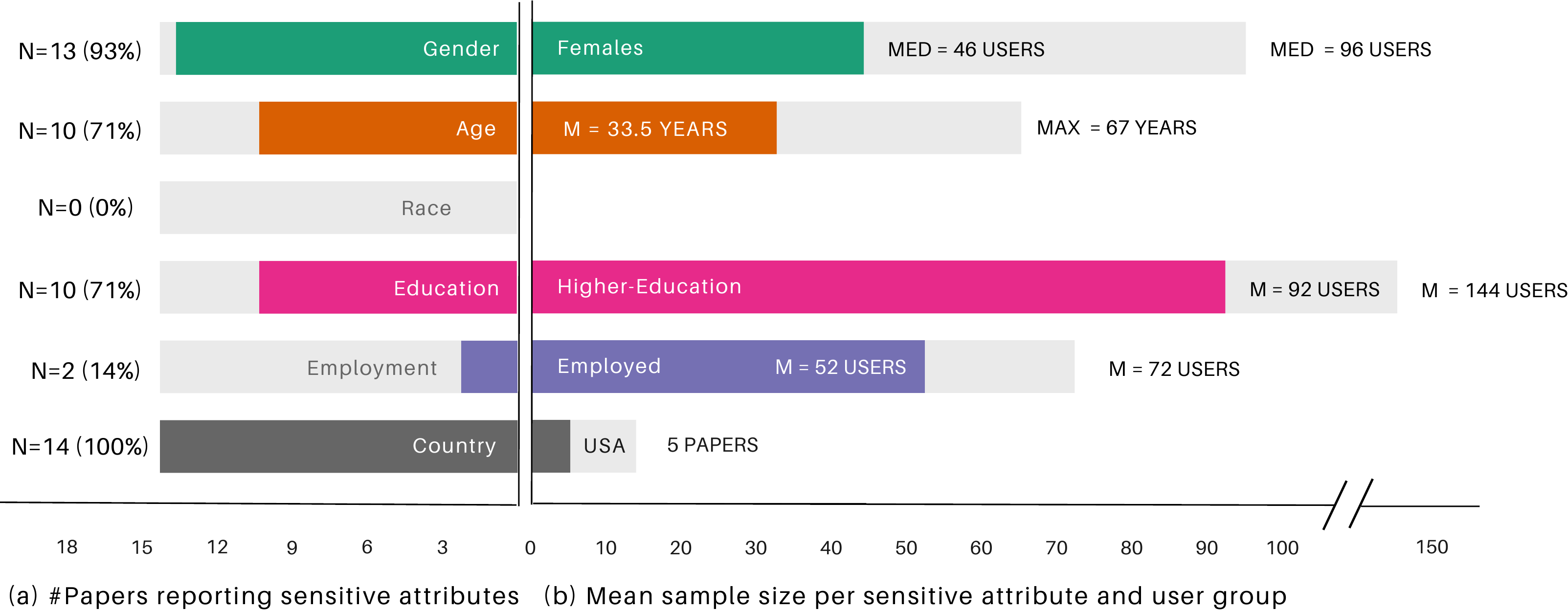}
  \caption{\textbf{Analysis of sensitive attributes and data size}. The bar plots show the percentage of papers reporting sensitive attributes (left) and the (mean) sample size in each paper subset (right). Demographics reporting is not standardized or comprehensive, with race and employment being the least reported ($0\%$ and $14\%$, respectively). While MobileHCI samples tend to be gender- and country-balanced, they remain WEIRD to some extent, as they consist of predominantly highly-educated and employed subjects.\label{fig:weird}}
  \Description{Two side-by-side bar plots showing the percentage of papers reporting certain sensitive attributes and the (mean) sample size in the subset of papers reporting that attribute. Specifically, 93\% report Gender, with a median of 96 users and 46 females. 71\% report Age, with the maximum mean age reported being 67 and the mean 33.5 years. 0\% report Race. 71\% report Education, with a mean of 144 users, 92 of which are highly educated, 14\% report Employment with a mean of 72 users, 52 of which are employed. Finally, 100\% report country of origin, with 5 out of 14 samples originating from the USA.}
\end{figure*}
\subsection{How WEIRD is MobileHCI?\label{WEIRDSection}}
Figure~\ref{fig:weird} shows the analysis of sensitive attributes in the included papers. The bar plot on the left (a) shows the percentage of papers reporting sensitive attributes: gender, age, race, education, employment, and sample country. The bar plot on the right (b) gives the mean sample size within each subset of papers reporting that sensitive attribute. In particular, gender was the most reported sensitive attribute. 93\% of included papers ($N=13$) disclosed this information. Within these papers, the median sample size was 96 users, and the median number of females was 46. Evidently, the community has stepped in the right direction to achieve more balanced, diverse, and representative datasets. Yet, there is plenty of room for improvement. Age and education were reported in 71\% of included papers ($N=10$). The mean sample age within these papers was 33.5 years old. For comparison, the maximum mean age reported was 67 years. In a world that is rapidly ageing, the MobileHCI community seems to be predominantly developing and testing on young populations. At the same time, within these 10 papers, the mean sample size was 144 users, and the mean number of highly-educated people was 92. Finally, none of the included papers mentioned the participants' race. Although these data are limited, they highlight not only that race is often overlooked in MobileHCI research (\S\ref{stateFairness}), but also that non-White populations are underrepresented in the datasets. There is a risk that models underperform for non-White users, but this may go unnoticed. On a positive note, out of the 14 papers for which the participants' country is reported or can be inferred, 5 engaged with US samples ($\sim$36\%), showing higher diversity than other UbiComp subcommunities \cite{yfantidou2023beyond}. 

\section{Discussion and Conclusion}
\label{sec:discussion}

The findings of our study reveal a gap in the integration of fairness considerations within the MobileHCI community. Only a small percentage of papers published between 2017 and 2022 explicitly addressed fairness (5\%), indicating a need for greater awareness and action in this area. While MobileHCI primarily focuses on understanding people, user experiences, and interactions, fairness principles are often overlooked. This highlights the importance of integrating fairness into the design and evaluation of mobile and wearable technologies, as algorithmic bias and discrimination can have significant ethical implications. \rev{Simiarly, prior work \cite{yfantidou2023beyond} highlights the absence of fairness considerations in IMWUT/UbiComp works --an adjacent venue--, where again only a small portion (5\%) of published papers adheres to bias reporting.} \rev{Drawing from these findings and borrowing from the ``Privacy by design'' literature \cite{cavoukian2009privacy}, we encourage a ``Fairness by design'' equivalent, requiring AI practitioners and researchers to consider fairness from the very beginning of any MobileHCI project or system design. This can be achieved by following fair practices across the ML lifecycle as introduced in prior work \cite{yfantidou2023beyond}.}

The prevalence of in-the-wild deployments in MobileHCI research is well-documented, providing insights into real-world user experiences. However, one noteworthy finding of this work is that biases can occur regardless of the deployment setting, emphasizing the need for fair systems across different contexts. Designing robust and fair systems that perform well beyond the lab is crucial to mitigate biases and ensure equitable outcomes for all users. The reviewed papers predominantly addressed gender and age as protected attributes, but it is essential to broaden the scope and consider other dimensions of diversity and potential sources of bias. Neglecting attributes such as race, ethnicity, socioeconomic status, and health conditions can perpetuate existing inequalities and biased outcomes. MobileHCI researchers should strive to incorporate and report a diverse range of attributes and input modalities in their studies to ensure fairness across multiple dimensions.

To promote algorithmic fairness in MobileHCI, we propose a few recommendations. Firstly, researchers should actively consider fairness as a core principle throughout the design and evaluation process of mobile and wearable systems. This involves integrating fairness metrics and bias mitigation strategies into the research workflow. \rev{For instance, there is a plethora of fairness toolkits, such as FairLearn \cite{bird2020fairlearn}, AIF360 \rev{\cite{bellamy2019ai}} and Aequitas \rev{\cite{saleiro2018aequitas}}, which provide diverse pre-processing, in-processing, and post-processing bias metrics and mitigation algorithms out-of-the-box --- some applicable regardless of input data types}. Additionally, the MobileHCI community should embrace interdisciplinary collaboration and draw insights from adjacent communities, including machine learning, fairness, accountability and transparency, human-computer interaction, and ubiquitous computing communities. By leveraging the expertise and methodologies from these fields, MobileHCI can advance the understanding and practice of algorithmic fairness.

Furthermore, researchers should actively pursue greater inclusivity and diversity in their studies, \rev{although this is contingent upon adequate funding}. This entails conducting studies with diverse participant populations \rev{(WEIRD)}, including individuals from different cultural backgrounds, age groups, and socioeconomic statuses. \rev{Beyond participants, researchers should strive for a diverse representation during data annotation. Considering that models encode label biases, they should be assessed by multiple people to ensure agreement but also strive for demographic diversity amongst them.} \rev{At the same time, researchers should validate collected data across different protected attributes to detect and address any anomalies, utilizing visualization tools, such as What-If \cite{wexler2019if} and FairLens \cite{panigutti2021fairlens}.} \rev{At the same time, the community itself could implement a mandatory data statement policy, requiring authors to report sensitive attributes concerning their participant samples. This builds on recent quests that advocate for data excellence~\cite{sambasivan2021everyone}, for example, by making data statements and datasheets for datasets mandatory for authors submitting their work.} By including underrepresented groups, MobileHCI can better understand and address potential biases and ensure the development of fair and equitable systems.

In conclusion, our study highlights the need for increased attention to algorithmic fairness in the MobileHCI community. While the current integration of fairness principles may be limited, MobileHCI has a unique opportunity to play a significant role in promoting fair and equitable interaction with mobile devices and applications. By prioritizing algorithmic fairness, embracing interdisciplinary collaboration, and promoting inclusivity and diversity, MobileHCI can contribute to the development of ethical and equitable mobile and wearable technologies that benefit all users.

\begin{acks}
This project has received funding from the European Union’s Horizon 2020 research and innovation programme under the Marie Skłodowska-Curie grant agreement No 813162. The content of this paper reflects only the authors' view and the Agency and the Commission are not responsible for any use that may be made of the information it contains. 
\end{acks}

\bibliographystyle{ACM-Reference-Format}
\bibliography{main}

\end{document}